\documentclass[a4paper,twocolumn,superscriptaddress,nofootinbib]{revtex4-1}
\usepackage[utf8x]{inputenc}
\usepackage{amsmath,amssymb,dsfont,mathtools,tensor}
\usepackage{hyperref,verbatim,color}

\DeclareMathOperator{\Tr}{Tr}
\numberwithin{equation}{section}

\begin{document}

\title{{\bf A Matrix Model for QCD}}

\author{\textbf{A. P. Balachandran}\email{balachandran38@gmail.com}}

\affiliation{Department of Physics, Syracuse University, Syracuse, N. Y. 13244-1130, USA}


\author{\textbf{Sachindeo Vaidya}\email{vaidya@cts.iisc.ernet.in}}

\affiliation{Centre for High Energy Physics, Indian Institute of Science, Bangalore, 560012, India}

\author{\textbf{Amilcar R. de Queiroz}\email{amilcarq@unb.br}}  

\affiliation{Instituto de Fisica, Universidade de
Brasilia, Caixa Postal 04455, 70919-970, Brasilia, DF, Brazil}

\affiliation{Departamento de F\'isica Te\'orica, Facultad de Ciencias, Universidad de Zaragoza, 50009 Zaragoza, Spain}

\begin{abstract}
Gribov's observation that global gauge fixing is impossible has led to suggestions that there may be a deep connection between gauge-fixing and confinement. We find an unexpected relation between the topological non-triviality of the gauge bundle and coloured states in $SU(N)$ Yang-Mills theory, and show that such states are necessarily impure. We approximate QCD by a rectangular matrix model that captures the essential topological features of the gauge bundle, and demonstrate the impure nature of coloured states explicitly. Our matrix model also allows the inclusion of the QCD $\theta$-term, as well as to perform explicit computations of low-lying glueball masses. This mass spectrum is gapped.

Since an impure state cannot evolve to a pure one by a unitary transformation, our result shows that the solution to the confinement problem in pure QCD is fundamentally quantum information-theoretic.

\end{abstract}

\maketitle

\section{Introduction}

That it is impossible to do global gauge fixing in QCD or in any non-Abelian gauge theory was first discussed by 
Gribov \cite{Gribov:1977mi}, for the specific case of Coulomb gauge. Singer \cite{Singer:1978dk} and 
Narasimhan and Ramadas \cite{Narasimhan:1979kf} proved this result rigorously for any choice of gauge 
by showing that the obstruction to global gauge fixing originates in the twisted nature of the gauge bundle on the 
QCD configuration space.

In this letter we construct a $(0+1)$-dimensional matrix model for gluons that captures this essential topological aspect of QCD. The proposed matrix model is a quantum mechanical model of $3\times (N^2-1)$ real matrices for $3+1$ dimensional $SU(N)$ Yang-Mills theory, and is free of the technical problems of QFT. Furthermore 
 for the case of $SU(2)$, this model may be argued to be equivalent to a certain $4$-body mechanical system.

Our model is inspired by Narasimhan and Ramadas' proof of the Gribov problem in the case of $SU(2)$. 
We here construct the Hamiltonian model for $SU(2)$. Another novelty is our adaptation of some crucial results 
of their work to higher $SU(N)$ and the corresponding Hamiltonian.

The proposed model brings many advantages for QCD calculations.  It is also suitable for the study of 't Hooft's 
large $N$ limit. As an explicit illustration of the power of our approach, we show that the gluon spectrum has a gap in our model. In lattice calculations this is taken as a signal for confinement. For $SU(2)$ and 
$SU(3)$, we here estimate the low-lying glueball masses using a simple variational method. We also show how to include the Chern-Simons (i.e. the well-known $\theta$) term.

Our most significant result stems from the fact the algebra of local observables in Yang-Mills theory is gauge-invariant: we find that coloured states 
of the theory are {\it necessarily} impure. For a non-Abelian gauge theory with local degrees of freedom, the proof is formal 
\cite{Balachandran:2014voa}, but in the matrix 
model approximation, this proof can be made rigourous. In pure QCD, the confinement problem is often informally stated as "what can't we see 
free gluons?". This result has deep implications for the confinement problem, as we shall see.

The matrix model solutions may be seen as a background, just as in a soliton model, around which one may 
quantize fluctuations. In this sense, the model contains a vacuum sector where the gauge potential is gauge 
equivalent to the zero field. We indicate how to construct multiparticle states for our gluon levels adapting 
standard techniques in soliton physics \cite{Balachandran:1991zj}.

Matrix models for Yang-Mills quantum mechanics have been suggested in the past, for instance in 
\cite{Savvidy:1982jk}, which has been explored by many researchers. However, their arguments for arriving at 
the model differ from ours, as does their potential.

Other investigations of Yang-Mills quantum mechanics involve approximating the gauge field by several 
$N \times N$ (unitary or hermitian) matrices. The potential in these models has interesting properties in the large 
$N$ limit, and several investigations have been carried out by 
\cite{O'Connor:2006wv, DelgadilloBlando:2007vx, DelgadilloBlando:2008vi, DelgadilloBlando:2012xg}. Again, 
these models differ from ours, since our model is based on a single rectangular $3\times(N^2-1)$ real matrix, with 
a kinetic energy term plus a potential originating from the $F_{ij}^2$ term.

\section{Matrix Model for \texorpdfstring{$SU(N)$}{Lg} gauge theory}

The topology of $(3+1)$-dimensional $SU(N)$ Yang-Mills gauge theory is well captured by the twisted fibre bundle ${\rm Ad}\,SU(N)\to \mathcal{M}_0\to \mathcal{M}_0/{\rm Ad}\,SU(N)$, where ${\rm Ad}\,SU(N)$ is the adjoint group of $SU(N)$, $\mathcal{M}_0$ are the $(N^2-1)\times (N^2-1)$ matrices of rank strictly greater than $(N-1)^2$. This last requirement assures that the action of ${\rm Ad}\,SU(N)$ on such $\mathcal{M}_0$ is free. These facts were worked out for the case of $SU(2)$ by Narasimhan and Ramadas. The generalization to higher rank is our construction \cite{Balachandran:2014voa}.

The matrix model construction starts from the general left-invariant one-form on $SU(N)$
\begin{equation}
	\label{Left-inv-1-form}
  \Omega=\Tr\left(\frac{\lambda_a}{2}~g^{-1}dg  \right) M_{ab} \lambda_b, \quad g\in SU(N),
\end{equation}
where $\lambda_a$, $a=1,...,N^2-1$, are the hermitian generators of the Lie algebra of $SU(N)$, $M$ is a real $(N^2-1)\times 
(N^2-1)$ matrix and $\Tr$ is in the fundamental representation of $SU(N)$. These $M$'s parametrize a 
submanifold  of the space of all connections $\mathcal{A}$, and the finite-dimensional bundle 
$\mathcal{M}_0/{\rm Ad}\,SU(N)$ captures the essential topology of the full gauge bundle. 

The rectangular nature of the matrix model emerges as follows. We fix a spatial-slice $S^3$ that can be isomorphically mapped to an $SU(2)$ embedded in $SU(N)$. The action of the left-invariant vector fields $X_i$, $i=1,2,3$ of this $SU(2) \subset SU(N)$ leads to
\begin{equation}
  \Omega(X_i)=-M_{ib}\frac{\lambda_b}{2}.
\label{connection}
\end{equation}
We next identify spatial vector fields with $iX_j$, $j=1,2,3$, so that the gauge field on the spatial-slice has components
\begin{equation}
  A_j=-iM_{jb}\frac{\lambda_b}{2}.
 \label{matrixgaugepot}
\end{equation}
The $SU(N)$ of colour acts on these gauge field components as
\begin{align}
  A_j\to h A_j h^{-1} \quad \text{ or } \quad M\to M({Ad}\,h)^T, \quad h\in SU(N). \nonumber
\label{gaugetr}
\end{align}
{\bf The Matrix Model Bundle is Twisted}: Let $\mathcal{M}$ be the space of real $(N^2-1)\times (N^2-1)$ matrices, and $\mathcal{M}_0$ be its subspace of matrices with rank strictly greater than $(N-1)^2$. As in \cite{Balachandran:2014voa}, we can show that ${\rm Ad} SU(N)R$ acts freely on $\mathcal{M}_0$.  Since $\mathcal{M}$ is contractible, then $\pi_j(\mathcal{M})=0$. Hence by Remark 3 to Theorem 6.2 in \cite{Narasimhan:1979kf}, $\pi_1(\mathcal{M}_0)=0$. This implies that $\mathcal{M}_0\neq \mathcal{M}_0/{Ad}\,SU(N)~\times~{Ad}\,SU(N)$ since $\pi_1 ({Ad}\,SU(N)) = \mathds{Z}_N$. Therefore ${Ad}\, SU(N)~\to~\mathcal{M}_0~\to~\mathcal{M}_0/{Ad}\,SU(N)$
captures the $SU(N)$ twist of the exact theory. For $N=2$, the twisted nature of $SO(3) \rightarrow \mathcal{M}_0 \rightarrow \mathcal{M}_0/SO(3)$, with ${\rm Ad}SU(2)=SO(3)$ is shown in the proof of Theorem 6.2 of \cite{Narasimhan:1979kf}. For $N=3$ an explicit proof can be found in \cite{Balachandran:2014voa}.

{\bf The Hamiltonian for $SU(N)$}: From the Yang-Mills action 
\begin{equation}
S=-\frac{1}{2g^2}\int d^4x \Tr F_{\mu \nu} F^{\mu\nu},
\label{YMaction}
\end{equation}
with $F_{\mu\nu}=\partial_\mu A_\nu -\partial_\nu A_\mu + [A_\mu, A_\nu]$, we obtain the Hamiltonian
\begin{equation}
H = \frac{1}{2}\int d^3 x \Tr \left(g^2E_i E_i - \frac{1}{g^2}F_{ij}^2 \right),
\label{hamiltonian}
\end{equation}
with $E_i$ being the chromoelectric field. For our matrix model, the canonical variables are $M_{i a}$ with the Legendre transformation of $\frac{d}{dt} M_{i a}$ as the corresponding momenta. The latter are identified with the matrix model chromoelectric fields $E_{i a}$. After quantization they satisfy $[M_{i a} , E_{j b} ]=i \delta_{ij} \delta_{a b}$.

From the vector potential $A_j$ (\ref{matrixgaugepot}), we obtain the associated curvature 
$F_{ij}=\left(d\Omega+\Omega\wedge \Omega\right)(iX_i,iX_j)$,
\begin{equation}
  F_{ij}=i\frac{\lambda_a}{2}\left(\epsilon_{ijk} M_{k a}  
  -  f_{abc} M_{ib} M_{jc} \right),
\label{2-form-field}
\end{equation}
where $f_{abc}$ are $SU(N)$ structure constants.

In the Hamiltonian (\ref{hamiltonian}), the potential $V(M)=-(\Tr F_{ij}F_{ij})/2g^2$ is written in the matrix model 
variables as  
\begin{align}
 V(M)&=\frac{1}{2g^2}\left( M_{k a} M_{k a} - \epsilon_{ijk} f_{abc} 
M_{i a} M_{j b} M_{k c} \right. \nonumber \\ 
&+ \left. \frac{1}{2} f_{a b c} f_{d e c} 
M_{i a} M_{j b} M_{i d} M_{j e}\right).
\end{align}
This manifestly invariant under gauge transformations $M\to M({Ad}\,h)^T, h\in SU(N)$.

The Hamiltonian is thus 
\begin{equation}
H=\frac{1}{R}\left(\frac{g^2 E_{i a} E_{i a}}{2} + V(M) \right),
\label{ham2}
\end{equation}
where we have inserted an overall length scale $R$ from dimensional considerations. As a quantum operator,  
\begin{equation}
H=-\frac{g^2 }{2R}\sum_{i, a} \frac{\partial^2}{\partial M_{i a}^2} + V(M).
\label{ham3}
\end{equation}
acts on the Hilbert space of functions $\psi(M)$ with scalar product
\begin{equation}
(\psi_1, \psi_2) = \int \Pi_{i, a} d M_{i a} \bar{\psi}_1 (M) \psi_2 (M).
\end{equation}

{\bf The \texorpdfstring{$SU(2)$}{Lg} Case}: Here $f_{abc} = 
f_{abc}$ and the $M$'s are $3 \times 3$ matrices. The potential simplifies to 
\begin{align}
V(M) &= \frac{1}{2g^2} \Big(\Tr M^T M -6 \det M  \nonumber \\
&+  \frac{1}{2} [(\Tr M^T M)^2 - \Tr M^T M M^T M] \Big).
\label{potential}
\end{align}
Furthermore, $V(M)$ is invariant under gauge transformations, $M \rightarrow M R^T$, with $R \in {\rm Ad}SU(2)\equiv SO(3)$.

The properties of the Hamiltonian (\ref{ham3}) are best fleshed out by performing a singular value decomposition (SVD) $M = R A S^T$, where $R$ and $S$ are $O(3)$ matrices, $A$ is a diagonal matrix with non-negative entries $a_i$ ordered as $a_1\geq a_2\geq a_3\geq 0$. The potential further simplifies to the quartic potential $2g^2 V(M) = (a_1^2 + a_2^2 + a_3^2) - 6 a_1 a_2 a_3 + (a_1^2 a_2^2 + a_1^2 a_3^2 + a_2^2 a_3^2)$.

$V(M)$ vanishes for $M=0$ and $M=R \in SO(3)$. For $R={\bf 1}$, these two configurations 
are gauge-related by a (large) gauge transformation of winding number $1$. Higher winding transformations are 
given by non-trivial higher winding number maps $SO(3) \rightarrow SO(3)$.

Symmetry of the equations under $a_i \leftrightarrow a_j$ suggests that all $a_i$ are equal at the extremum. It is thus enough to compute 
$\partial V/\partial a_1 = (1/g^2)(a_1 -3a_2 a_3 + a_1(a_2^2+a_3^2)=0$, and set $a_1=a_2=a_3\equiv a$ which gives $a=0, 1/2, 1$ as extrema. The Hessian $[\partial^2 V/\partial a_i \partial a_j]$ is  positive definite at both $a=0$ and $a=1$. At $a=0$ all its eigenvalues are degenerate, while at $a=1$ only two of its eigenvalues are degenerate. This is so although these are related by a large gauge transformation. The physical 
consequences of this is unclear to us. The extremum at $a=1/2$ is a saddle point, with one negative eigenvalue for the Hessian.

It is interesting that on performing the SVD of $M$, the kinetic term of our Hamiltonian becomes closely 
related to that of a $4$-particle quasi-rigid body \cite{zickendraht,iwai}.

{\bf The \texorpdfstring{$SU(3)$}{Lg} Case}: Now the $M$'s are $3 \times 8$ matrices. Rather than perform the SVD of $M$, we will work directly with the variables $M_{ia}$ and the Hamiltonian (\ref{ham3}) to find the spectrum by a variational calculation.


The potential $V(M)$ in (\ref{ham3}) grows at least quadratically in $M_{ia}$ as $|M_{ia}| \rightarrow \infty$ and is smooth everywhere. This suggests that the spectrum is gapped. For small values of $|M_{ia}|$ the anharmonic terms are small. However, we cannot treat these terms 
 as a perturbation of the harmonic oscillator, since it is notoriously singular  \cite{Loeffel:1970fe,Graffi:1990pe,Mathews:1977rf}. We sidestep this difficulty 
  by using the variational method to estimate the energy levels.

Let $H_0$ be the Hamiltonian without the anharmonic term. The eigenfunctions of $H_0$ are of the form $f(M_{i a}) e^{-M_{i a} M_{i a}/2g^2}$, where $f(M_{i a})$ are products of Hermite polynomials in $3(N^2-1)$ variables $M_{i a}$. We take the variational ansatz for the ground state to be $\Psi^0_\beta = A_0 e^{-\frac{\beta}{2g^2} M_{i a} M_{i a}}$, $A_0 = \left(\frac{\beta}{\pi g^2}\right)^{\frac{3}{4}(N^2-1)}$, with variational parameter $\beta$. 
Minimizing $\langle \Psi^0_\beta |H |\Psi^0_\beta \rangle$ w.r.t $\beta$ gives us the variational ground state energy $E^{0}_\text{var}(g)$, which we have plotted in Fig 1 as a function of t'Hooft coupling $t=g^2 N$.

Similarly, for the first excited state, we take as the variational ansatz $\Psi^1_{\beta,a} = A_1 (\beta,g)M_{i a} e^{-\frac{\beta}{2g^2} M_{i a} M_{i a}}$.
This state carries a color index $a$. The variational energy of this state is plotted in figure 2.


Both $\Psi^0_\beta$ and $\Psi^1_{\beta,a}$ are insensitive to the cubic term in the Hamiltonian. The simplest ansatz with this sensitivity is 
$\Phi^1_{(\beta,\gamma)} = B_1 (M_{i a} +\gamma \,\epsilon_{ijk} f_{abc} M_{j b} M_{k c}) e^{-\frac{\beta}{2g^2} M_{i a} M_{i a}}, \gamma \in \mathds{C}$. This has three variational parameters: $\gamma,\gamma^*$ and $\beta$. Minimizing $\langle \Phi^1_{(\beta,\gamma)} |H |\Phi^1_{(\beta,\gamma)} \rangle$ w.r.t $\beta,\gamma,\gamma^*$ gives the energy of this state, which is plotted in Fig 3.

\begin{figure}[h]
        \centerline{
               \mbox{\includegraphics*[width=3in]{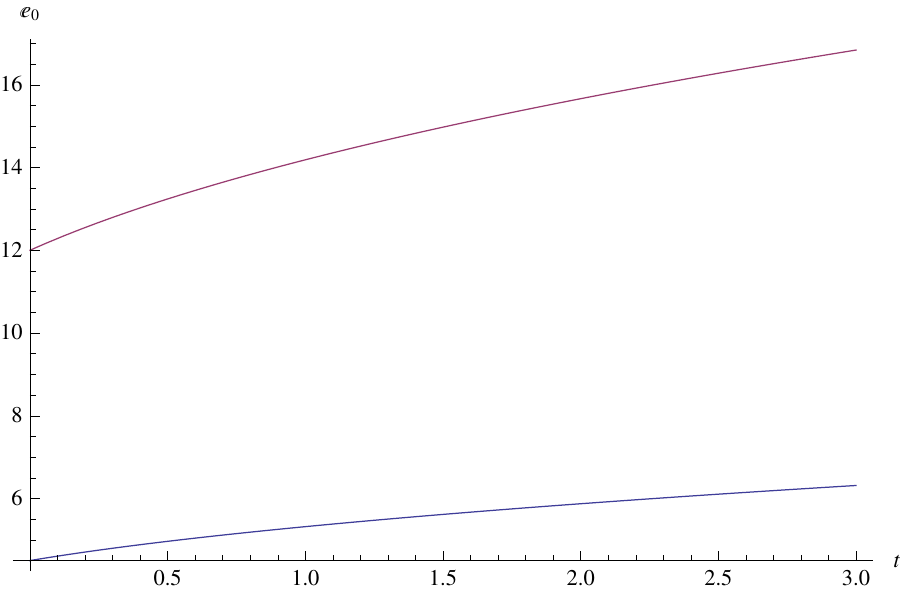}}
                             }
\caption{The blue line is for $N=2$, the red line is for $N=3$. The energy is in units of $1/R$ and $t=g^2N$.}
\end{figure}

%

\begin{figure}[h]
        \centerline{
               \mbox{\includegraphics*[angle=0,width=3in]{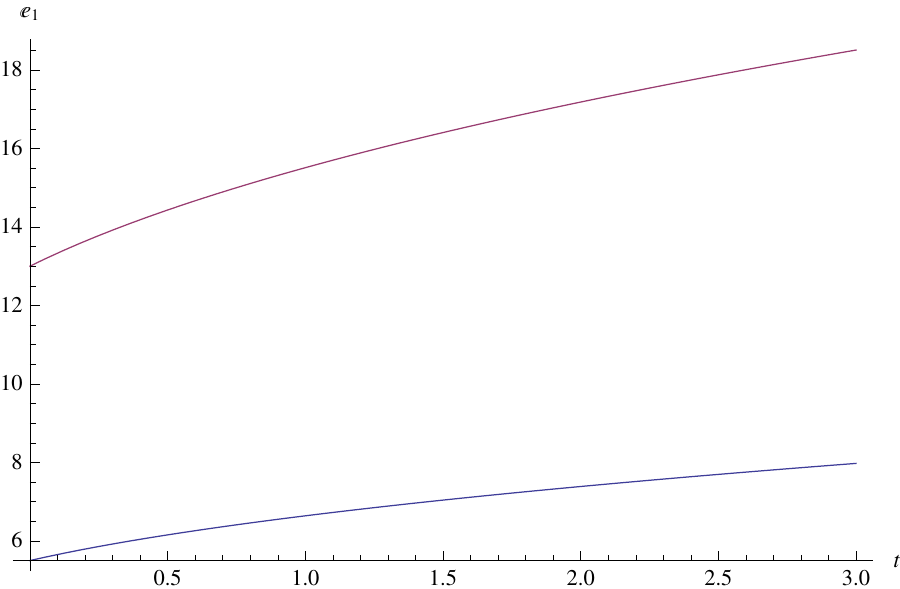}}
                             }
\caption{The blue line is for $N=2$, the red line is for $N=3$. The energy is in units of $1/R$ and $t=g^2N$.}
\end{figure}

\begin{figure}[h]
        \centerline{
               \mbox{\includegraphics*[angle=0,width=3in]{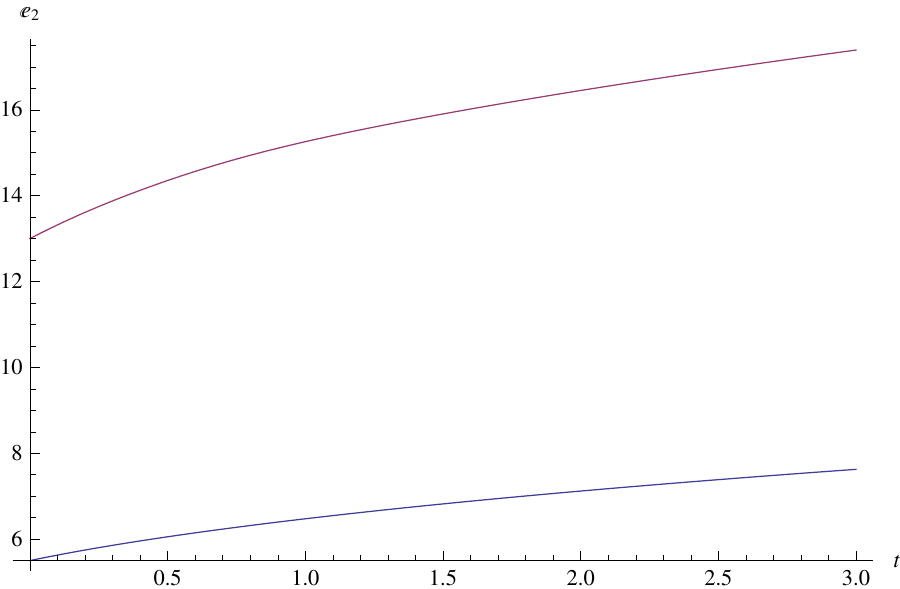}}
                             }
\caption{The blue line is for $N=2$, the red line is for $N=3$. The energy is in units of $1/R$ and $t=g^2N$.}
\end{figure}

Our variational energy estimate is rather crude, and we expect that the variational 
estimate differs significantly from the true energy  for large values of the coupling $t$. Much better numerical estimates may be obtained by taking more sophisticated variational ansatz for the wavefunctions. 

\section{Coloured States are Mixed}

The algebra of observables of the matrix model consists of all polynomials in $M_{ia}$ and $E_{ia}$ subject to the relation $\left[M_{ia},E_{jb} \right]=i\delta_{ij}\delta_{ab}$. Furthermore, the observables have to be singlets under gauge transformations because they commute with Gauss' law. On the other hand, a state may have color index. For instance, we may consider a color state $|\cdot,a\rangle$, $a=1,...,N^2-1$. Note that the projector $\mathds{P}=\sum_a |\cdot,a\rangle\langle\cdot,a|$ is a color singlet, but not its individual terms.

We can now proceed with the reconstruction of the so-called GNS Hilbert space, which shows that coloured states are impure \cite{Balachandran:2012pa,Balachandran:2013kia}. Indeed, it is enough to consider the preparation of a state where the value of projector $\mathds{P}$ is $1$. The fact that we are restricting the value of $\mathds{P}$ to be $1$ means that any combination $\sum_a \alpha_a|\cdot,a\rangle$, with $|\cdot,a\rangle$ being normalized and $\sum_a |\alpha_a|^2\equiv \sum_a \mu_a=1$ is an allowed outcome. This state gives on an observable $K$ the expectation value 
\begin{equation}
\label{state-on-K-1}
  \sum_a \mu_a \langle \cdot,a |K|\cdot,a\rangle.   
\end{equation}
Recall that $K$ is a colour singlet, so that  $\langle \cdot,a |K|\cdot,a\rangle$ is independent of $a$ and therefore (\ref{state-on-K-1}) is independent of the choice of $\mu_a$. Hence the observation of $1$ for $\mathds{P}$ in the coloured sector leads to a mixed state. As there is no coloured observable, one cannot observe $|\cdot,a\rangle\langle\cdot,a|$ individually. Equivalently, we cannot prepare a pure coloured state.

There are many analogies to this situation elsewhere. For example, in scattering theory, if the in-state is a pure spin state of spinning particles, so is the final out-state. But if experiments do not detect the out-state spin, we sum the cross-sections (and not the amplitude) over out-state spins. The mean value of an observable $K$ which commutes with spin is also given by an expression like \ref{state-on-K-1}) where $\sum_{a} \alpha_a |\cdot,a\rangle $ is the out-state, and $a$ denotes spin components. Hence on $K$, the out-state is mixed.

\section{Final Remarks}

The $C^*$-algebra $\mathcal{C}(\mathcal{M})$ of the observables contains colour singlet functions of $M$. The full 
$\mathcal{C}(\mathcal{M})$ is generated by such operators. The straightforward adaptation of section 3 shows that coloured states restricted to $\mathcal{C}(M)$ are not pure. This fact will affect correlators and partition 
functions, and hence physical predictions. In particular, our reduced matrix model for pure QCD gives a gapped spectrum and discrete levels for glueballs. Generalisation to other non-abelian gauge groups is straightforward. 

Further remarks follow: {\bf (i)} $A_i$ can couple with quarks via the covariant derivative $\nabla_i=\partial_i+A_i$ in the Dirac operators. This is their only modification in the $A_0=0$ gauge; {\bf (ii)} QCD $\theta$-states $|\cdot,\theta\rangle$ are constructed with a Chern-Simons $3$-form action which, on using (\ref{matrixgaugepot}) and (\ref{2-form-field}), is seen to be given in the reduced matrix model, on using (\ref{matrixgaugepot}) and (\ref{2-form-field}), by $S_{\rm CS}(M)=\frac{1}{4} \Big[ \Tr(M^T M) +\frac{1}{6} \epsilon_{ijk} f_{abc} M_{ia}M_{jb}M_{kc}\Big] $. The overall $1/4$ is fixed by requiring that for a pure gauge, where $M=\bf{1}_{3\times 3} \oplus \bf{0}_{5\times 5}$, where $\bf{1}_{3\times 3}$ is in the $SU(2)$ subspace, the $S_{\rm CS}$ becomes the winding number $1$. Then under a gauge transformation $A~\to~hAh^{-1}+hdh^{-1}$,  $S_{\rm CS}(A)$ changes by the winding number $N(h)$ of the map $h$ \cite{Balachandran:1991zj,hoppe}, $S_{\rm CS}(hAh^{-1}+hdh^{-1})=N(h)+S_{\rm CS}(A), \quad N(h) \in \mathds{Z}$. Hence $e^{i\theta S_{\rm CS}(hAh^{-1}+hdh^{-1})}=e^{i\theta N(h)}e^{i\theta S_{\rm CS}(A)}$ and
\begin{equation}
  |\cdot,\theta\rangle = e^{i\frac{\theta}{2\pi} S_{\rm CS}} |\cdot,0\rangle;
\end{equation}
{\bf (iii)}  We can build multiparticle states for our gluon levels from (\ref{Left-inv-1-form}) by changing $u(\vec{x})$ to higher winding number maps as in Skyrmion physics \cite{Balachandran:1991zj}; {\bf (iv)} That coloured states are impure states have deep implications for the confinement problem. Consider the time-evolution of a pure (and hence colourless) state. Since time evolution in quantum theory is given by a unitary operator, this state will {\it never} evolve to a coloured state. Thus it is impossible to create a free gluon starting from a colourless state by any Hamiltonian evolution, and in particular by scattering.

\section{Acknowledgements}
We are very grateful to M. S. Narasimhan for many discussions and inputs. We have also benefitted from the suggestions of Sanatan Digal, Denjoe O'Connor and Apoorva Patel. Manolo Asorey and Juan Manuel Perez-Pardo have explained to us specific issues connected to domains of operators. APB thanks the group at the Centre for High Energy Physics, IISc, Bangalore, and especially Sachin Vaidya for hospitality. AQ thanks the DFT of the Universidad de Zaragoza for the hospitality and nice atmosphere. In particular, AQ thanks Monolo Asorey for fruitful discussions.  AQ is supported by CAPES process number BEX 8713/13-8.

\end{document}